# X-Cross: Image Encryption Featuring Novel Dual-Layer Block Permutation and Dynamic Substitution Techniques

Hansa Ahsan[1], Safee Ullah[1], Jawad Ahmad[2*], Aizaz Ahmad Khattak[3], Muhammad Ali[1], Muhammad Shahbaz Khan[3*]

[1]Department of Electrical Engineering, HITEC University, Taxila, 47080, Pakistan.
[2*]Cyber Security Center, Prince Mohammad Bin Fahd University, Al Khobar, 31952, Saudi Arabia.
[3*]School of Computing, Engineering and the Built Environment, Edinburgh Napier University, Edinburgh, EH10 5DT, UK.

*Corresponding author(s). E-mail(s): jahmad@pmu.edu.sa; muhammadshahbaz.khan@napier.ac.uk;
Contributing authors: 20-ee-073@student.hitecuni.edu.pk; safee.ullah@hitecuni.edu.pk; 40614576@live.napier.ac.uk; 20-ee-026@student.hitecuni.edu.pk;

**Abstract**

In this digital age, ensuring the security of digital data, especially the image data is critically important. Image encryption plays an important role in securing the online transmission/storage of images from unauthorized access. In this regard, this paper presents a novel diffusion-confusion- based image encryption algorithm named as X-CROSS. The diffusion phase involves a dual-layer block permutation. It involves a bit-level permutation termed Inter-Bit Transference (IBT) using a Bit-Extraction key, and pixel permutation with a unique X-cross-permutation algorithm to effectively scramble the pixels within an image. The proposed algorithm utilizes a resilient 2D chaotic map with non-linear dynamical behavior, assisting in generating complex Extraction Keys. After the permutation phase, the confusion phase proceeds with a dynamic substitution technique on the permuted images, establishing the final encryption layer. This combination of novel permutation and confusion results in the removal of the image's inherent patterns and increases its resistance to cyber-attacks. The close to ideal statistical security results for information entropy, correlation, homogeneity, contrast, and energy validate the proposed scheme's effectiveness in hiding the information within the image.

# 1 Introduction

With the rapid advancement in network technology, the transmission of digital data, especially images, has increased across different networks [1]. Information centers in fields such as healthcare, military operations, education, and finance now frequently exchange sensitive information through networks that are prone to various cyberattacks [2]. Therefore, safeguarding this digital information is essential to prevent the unauthorized access of sensitive data and individual privacy [3]. Image encryption techniques play a vital role in protecting the image data from unauthorized access [4]. Image encryption involves the transformation of original data (plain image) into a concealed form (cipher image) with the help of secret keys [5]. Traditional cryptographic ciphers like AES (advanced encryption standards), DES (data encryption standards), 3DES, and the International Data Encryption Algorithm (IDEA) are primarily designed for text information and don't perform well for images that have high pixel correlation [6].

According to Claude Shannon, an image encryption algorithm should involve two operations: confusion and diffusion [7]. Confusion is usually defined as the process of changing pixel values using a key, usually by substituting one value with another, also called substitution, whereas diffusion is defined as the process of changing the position of the pixels in the image [8, 9]. Various image encryption schemes can be found in literature that utilize the aforementioned two processes [10].

Recently, chaos theory is being utilized in image encryption to make the algorithms more unpredictable and secure. Chaotic maps generate highly non-linear pseudo-random sequences and help in generating encryption keys [11]. Chaotic maps are highly sensitive to initial conditions and possess inherent ergodicity and unpredictable behavior [12]. Chaotic maps generate unpredictable patterns that are used in the permutation/diffusion phase [13]. Apart from being utilized in traditional encryption scheme, chaotic maps are also being utilized in quantum image encryption [14].

The strength of the confusion and diffusion module is critically important when developing any encryption method. A good diffusion/permutation block should effectively break pixel-to-pixel correlation [15]. Similarly, a good confusion module should also be lightweight in addition to being highly non-linear. Keeping these characteristics in mind, this paper outlines a novel approach to image encryption technique, featuring a novel key generation module, a novel block-permutation module, and a modified dynamic substitution module. The main contributions of this paper are:

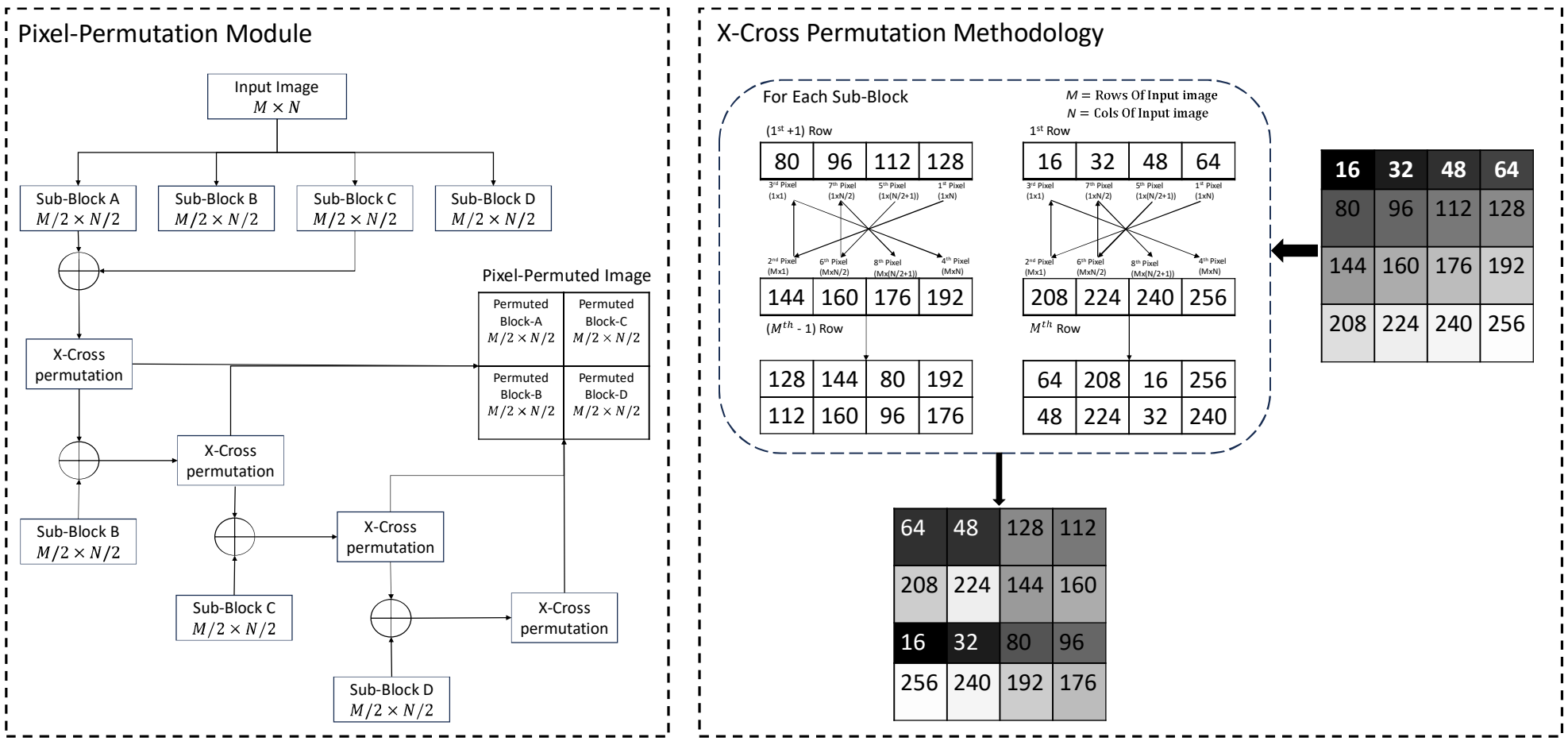

**Fig. 1**: X-Cross Permutation Methodology and Pixel Permutation Module

1. This paper introduces a novel chaos-based key generation technique that generates extraction keys. It utilizes hyperchaotic maps to produce random arrays that undergo sorting to generate highly sensitive extraction keys.
2. This paper introduces a novel dual-layer permutation module. The first layer performs permutation at bit level by utilizing a chaotic extraction key and block XOR operations. The second layer is the X-CROSS (pixel permutation) technique, which operates at pixel level to ensure effective scrambling. This dual-layer permutation effectively breaks correlation by spreading pixels throughout the image. The permutation module is given in Figure 1.

# 2 Secret Key Generation

Before explaining the proposed image encryption scheme, it is important to elaborate on some foundational components, such as the key generation algorithms utilized in the scheme. These include extraction keys generated through hyperchaotic maps and generation of an operation selection matrix for the confusion protocol.

## 2.1 Generation of Extraction Key

### 2.1.1 2D Logistic-Sine Hénon Map

The 2D-LSHM is a hyperchaotic map that merges the features of the Logistic and Sine maps with the Hénon map. These maps are used to generate pseudorandom numbers that are employed in various cryptographic applications. The equation of LSHM is defined by:

$$x_{n+1} = k_1 \left[1 + \alpha coscos\, (\pi\ x_n)^{\beta}\ \right] mod 1 \tag{1}$$

$$y_{n+1} = k_2 \left(coscos y_n \left(1 - x_n\right)\right) (2) \tag{2}$$

The parameters k1, k2, $\propto$ and $\beta$ determine the behavior of map. The proposed diffusion scheme generates two permutation keys by iterating these two equations.

#### 2.1.2 Random Extraction Array

To generate the random extraction array (REA) with dimensions $1 : M/2 \times N/2$ (such as $1 : (128 \times 128) \times 8$), we iterate the equations (1) by $(M/2 \times N/2) \times 8$ times because this implies to inter-bit transference, and each pixel contains 8-bit. After each iteration, the resultant arrays contain random numbers in fractional form. To handle these fractions, we scale each number and round them to obtain integers with finite precision sequences, which denoted as $REA_1'$.

$$REA_1' = round\left(R_1 \times 10^5\right) \tag{3}$$

However, a modulus operation is applied to ensure that the rounded random arrays contain values within the range of 1 to 256. This operation helps us limit the values to the desired range.

$$REA = \ mod\left(REA_1', 256\right) \tag{4}$$

Through these steps, two random arrays $R_1$ and $R_2$ are generated using 2D LSHM map for Extraction-key1 and Extraction-key2

Subsequently, to facilitate the construction of extraction keys, the arrays undergo sorting in ascending order, and the corresponding sorted indices are extracted to form an array structure. This array serves as the initial template for the extraction key. The extraction key generation process is iterated twice, yielding four distinct extraction keys denoted as Extraction-key1 and Extraction-key2. The remaining Extraction-key3 and Extraction-key4 are inverted form of Extraction-key1 and Extraction-key2. These keys collectively contribute to the diffusion mechanism. Figure 2 provides an illustration of how the plain image block is defused via Extraction keys.

### 2.2 Generation of Operation Selection Key

#### 2.2.1 Combined logistic-tent (CLT) map

An operation selection key is required to select the s-box and substitute the selected s-box value. For this purpose, CLT map is utilized. Mathematically, the Combined logistic-tent map is defined as follows:

$$Z_{i+1} = \begin{cases} mod\left(\lambda * Z_i * (1 - Z_i) + \frac{\alpha * Z_i}{2}, 1\right); & \text{for } Z_i \\ mod\left(\lambda * Z_i * (1 - Z_i) + \frac{\alpha * (1 - Z_i)}{2}, 1\right); & \text{for } Z_i \geq 0.5 \end{cases} \tag{5}$$

where $'\lambda'$ and $' \propto '$ are control parameters of ranges (3.5, 4) and (2, 4), respectively. $Z_{i+1}$ and $Z_i$ represent the outputs of the map within the range (0, 1) for $(i+1)^{th}$ and $i^{th}$ iterations, respectively. This map exhibits chaotic behavior across the entire range of (0, 4).

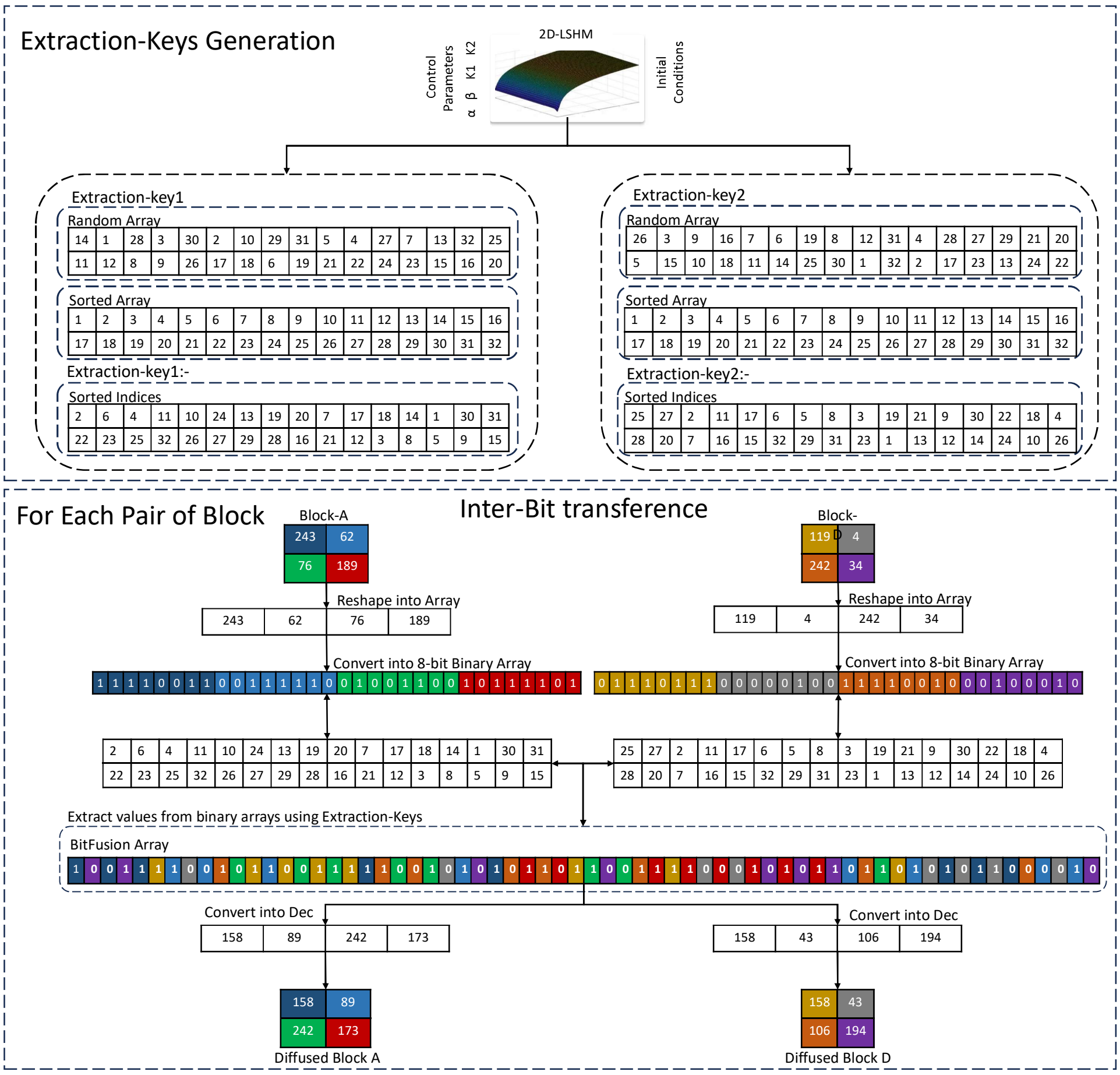

**Fig. 2**: Extraction-keys Generation and Inter-Bit transference Process

### 2.2.2 Operation Selection Matrix

To generate a random matrix for operation selection O with dimensions $M \times N$ (such as 256×256), we iterate equations (5) $M/2 \times N/2$ times. After each iteration, the resultant matrix contains random numbers in fractional form. To handle these fractions, we scale each number and round them to obtain integers with finite precision sequences, which denoted as O'.

$$O' = round\left(O \times 10^3\right) \tag{6}$$

However, to ensure that the random arrays contain values 0,1 and 2, a modulus operation is applied.

$$O' = mod(O', 3) \tag{7}$$

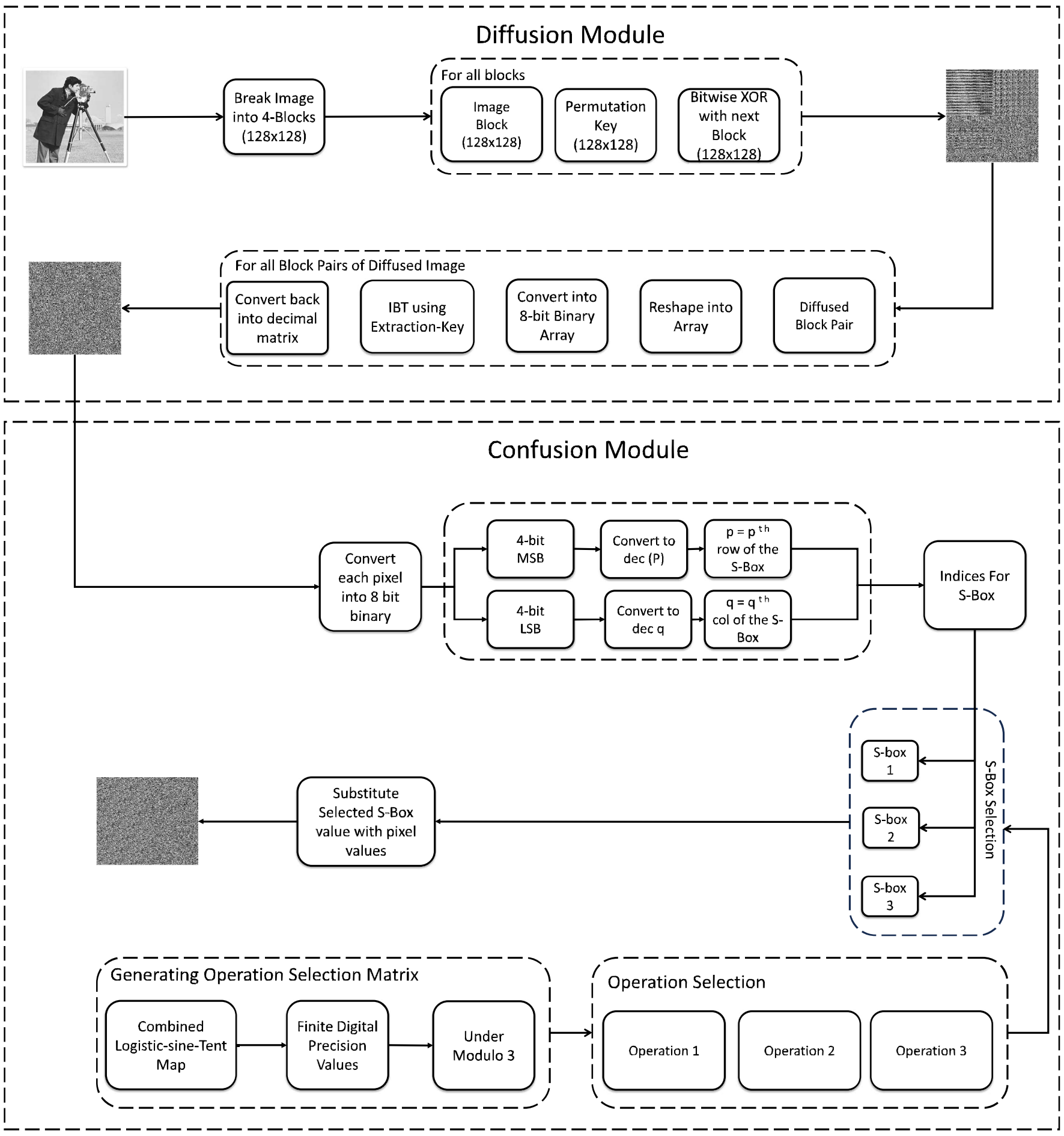

**Fig. 3**: The Proposed Encryption Scheme

# 3 The Proposed Encryption Algorithm

This paper introduces a novel diffusion and confusion-based encryption algorithm that consists of following components and encryption modules as depicted in Figure 3.

## 3.1 Novel block Permutation Technique

This paper introduces dual-layer block permutation technique. This technique involves pixel level permutation and bit-level permutation.

### 3.1.1 X-CROSS

The core component of this proposed scheme is the X-Cross permutation, which involves extracting pixels in a cross manner to enhance the encryption process as depicted in Figure 1. The image is initially decomposed into four equal size blocks of size $M/2\times\ N/2$ where $M/2\times\ N/2$ is $(128\times128)$. After decomposing the image, one of the four equal sized blocks is selected, then the 1st and last rows of a block are selected to extract pixel values in X-Cross manner. The sequence of X-Cross for pixel extraction is as follows:

1. Extraction of the 1st pixel from $(1\times N/2)$ position.
2. Extraction of the 2nd pixel from $(M/2\times 1)$ position.
3. Extraction of the 3rd pixel from $(1\times 1)$ position.
4. Extraction of the 4th pixel from $(M/2\times N/2)$ position
5. Extraction of the 5th pixel from $\left(1\times\left(\frac{N}{4}\right)+1\right)$ position.
6. Extraction of the 6th pixel from $(M/2\times N/4)$ position.
7. Extraction of the 7th pixel from $(1\times N/4)$ position.
8. Extraction of the 8th pixel from $\left(M/2\times\left(\frac{N}{4}\right)+1\right)$ position.

This is how all blocks are permuted. Besides, the blocks A and C are bit XORed with each other prior to applying X-Cross permutation. This XORed matrix then undergoes X-Cross, which is then Bit-XORed with the Block B. This process contues for all blocks as depicted in Figure 1.

#### 3.1.2 Inter-Bit Transference— IBT

This process begins by selecting two blocks of dimensions $M/2\times\ N/2$from the permuted image. Each block is converted into a linear array of size $1:M/2\times\ N/2$. Then each pixel is transformed into 8-bit binary resulting in an array of size $1:(M/2\times N/2)\times 8$.

By utilizing the Extraction-keys, the bits from the block arrays are extracted randomly. These extracted bits are placed into a newly named array called ”BitFusionArray”. This procedure completes the inter-bit transference.

### 3.2 Dynamic Substitution

In the proposed encryption process, an operation selection based dynamic substitution technique developed by M.S. Khan et al. in [8] is utilised. Three S-boxes are used for the substitution process, which are selected randomly based on the chaotic sequence generated by the chaotic map. The catch here is that, each value that is selected from the S-box first undergoes an additional operation before replacing the pixel value in the image. This ensures optimum randomness and improved confusion. The process is described in Figure 3.

## 4 Results and Security Analysis

This section presents the outcomes of the proposed encryption scheme and includes a comprehensive statistical security analysis to assess its effectiveness. Figures 4 and 5 illustrate the visual outcomes of the encryption process, highlighting the successful concealment of information within the input image.

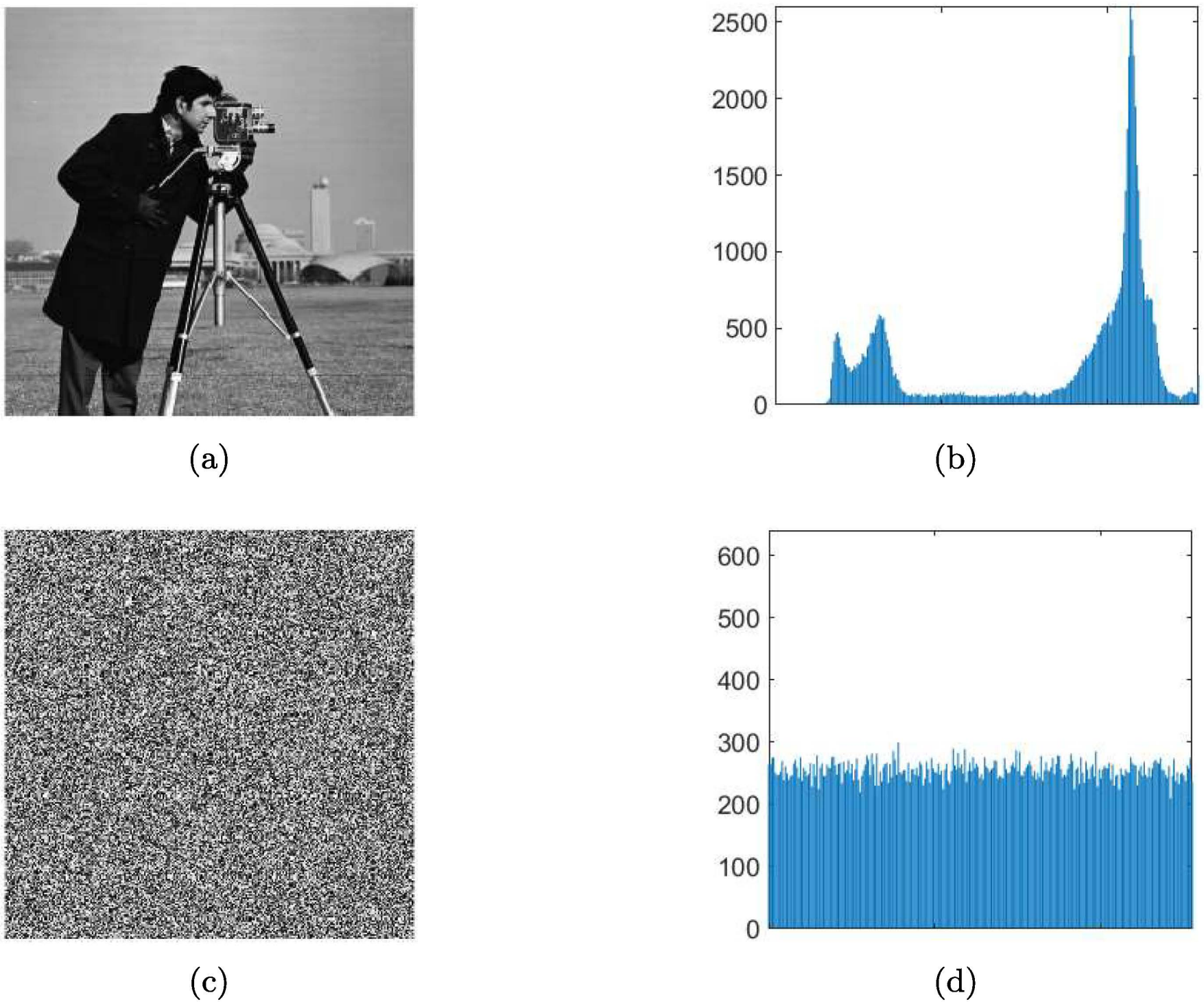

**Fig. 4**: Entropy of (a-b). Plain text Image, (c-d). Encrypted Image.

## 4.1 Correlation Coefficients Analysis

Correlation coefficients measure the similarity between different pixels in an image. A low correlation signifies high randomness, effectively removing pixel patterns from the original image. As demonstrated in Fig.5. The correlation coefficients are well-dispersed, and as shown in Table 1, the correlation is close to zero. This indicates the high effectiveness of the proposed scheme.

**Table 1**: Security Analysis

| | | |
|---|---|---|
| **Entropy Analysis** | Entropy | 7.9978 |
| **GLCM** | Correlation | 0.0000658 |
| | Contrast | 10.948 |
| | Energy | 0.0030 |
| | Homogeneity | 0.36417 |

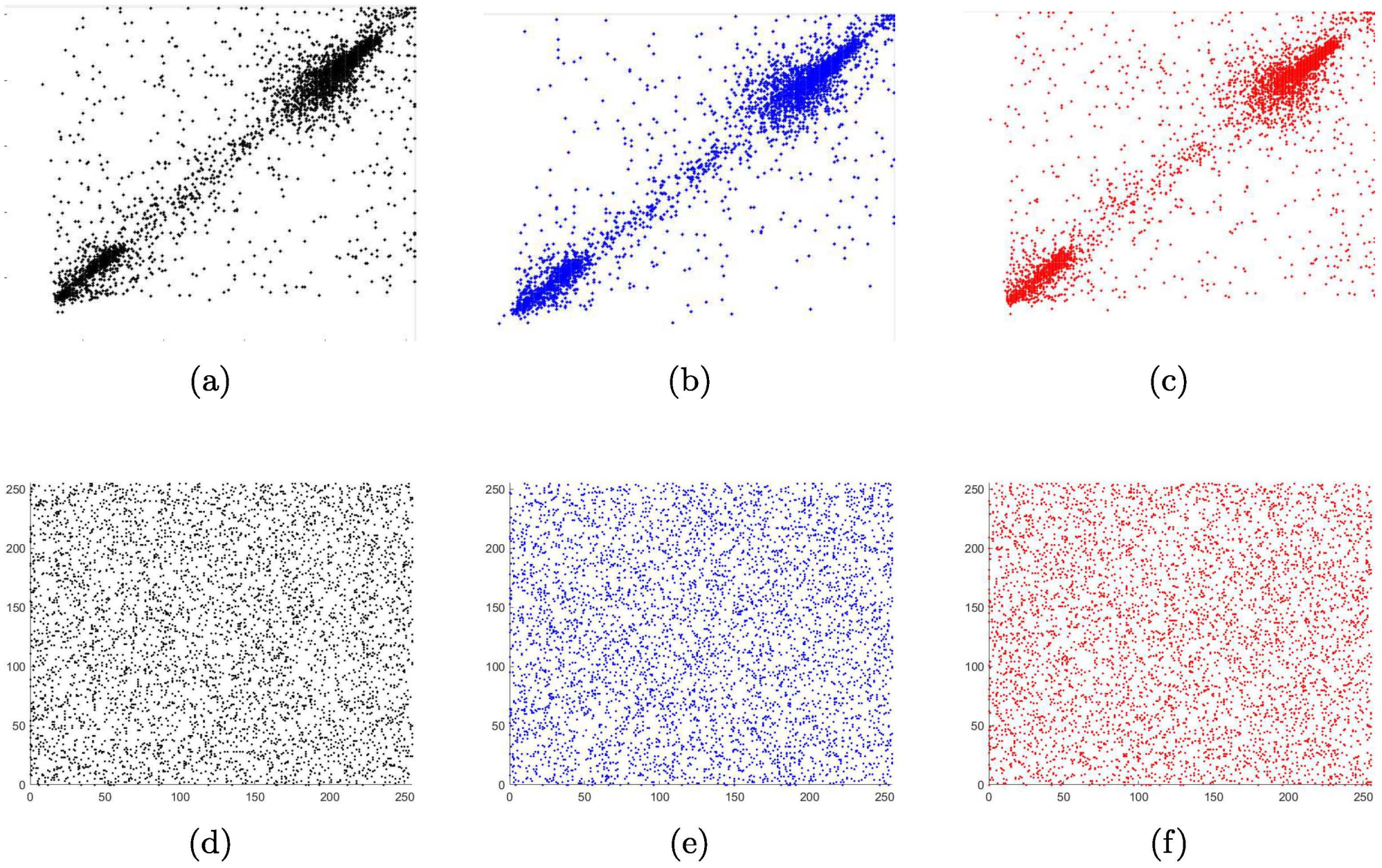

**Fig. 5**: .Correlation Coefficient Analysis of X-Cross (a-d).Horizontal Coefficients, (b-e).Vertical Coefficients, (c-f).Diagonal Coefficients.

## 4.2 Information Entropy Analysis

Information entropy provides a measure of the disorder and uncertainty of data within an image. For a grayscale image the ideal entropy should be 8. The results of information entropy are given in Table 1. Results show that the entropy of the cipher image is ideally close to 8.

## 4.3 Histogram Analysis

The histogram illustrates the distribution of pixel values across various gray levels in an image. For an image to be considered effectively encrypted, its histogram should be uniformly distributed. The uniformly distributed histogram of the cipher image shown in Fig.4.(d). demonstrates the effectiveness of the proposed encryption scheme.

## 4.4 GLCM-Parameters

Gray-level co-occurrence matrix (GLCM) parameters provide crucial information about the structure of an image. These parameters typically include contrast, correlation, energy, and homogeneity. For a secure ciphertext image, energy, homogeneity, and correlation value should be low, while contrast should be high. This pattern is evident in the results shown in Table 1.

# 5 Conclusion

The X-CROSS encryption scheme introduces a robust method for securing image data through a combination of bit-level and pixel-level permutations, effectively breaking pixel correlations. By utilizing hyperchaotic maps for key generation, the scheme ensures the creation of highly unpredictable and complex keys. These keys are employed in a novel dual-layer permutation process, which includes Inter-Bit Transference (IBT) and X-CROSS pixel permutation. This approach diffuses the image data and enhances security. Additionally, a dynamic substitution process is applied to further obscure the image data, removing inherent patterns and increasing resistance to cyber-attacks. Extensive experimental results validate the scheme's effectiveness, demonstrating high entropy, low correlation, and ideal values for homogeneity, contrast, and energy. Overall, X-CROSS offers a highly secure and efficient solution for image encryption, incorporating novel techniques for improved performance and security. Future work may include moving towards quantum cryptography for image encryption.